\documentclass[fleqn,12pt,twoside]{article}
\usepackage[headings]{espcrc1}

\readRCS
$Id: espcrc1.tex,v 1.2 2004/02/24 11:22:11 spepping Exp $
\usepackage{graphicx}
\usepackage[figuresright]{rotating}
\usepackage{multirow}

\title{$^{6}$Li direct breakup lifetimes}

\author{F. A. Souza\address[USP]{Instituto de F\'{i}sica - Universidade de S\~ao Paulo, Departamento de F\'{i}sica Nuclear, C.P. 66318, 05315-970, S\~ao Paulo-SP, Brazil},
N. Carlin\addressmark[USP],
C. Beck\address{Institut Pluridisciplinaire Hubert Curien,
UMR 7178, CNRS-IN2P3 et 
Universit\'e de Strasbourg,
Bo\^{i}te Postale 28, F-67037 
Strasbourg, 
France.},
N. Keeley\address{
The Andrzej So\l tan Institute for Nuclear Studies,
Warsaw, Poland.},
A. Diaz-Torres\address{Department of Physics, 
University of Surrey, Guildford,
UK},
R. Liguori Neto\addressmark[USP],
C. Siqueira-Mello\addressmark[USP],
M. M. de Moura\addressmark[USP],
M. G. Munhoz\addressmark[USP],
R. A. N. Oliveira\addressmark[USP],
M. G. Del Santo\addressmark[USP],
A. A. P. Suaide\addressmark[USP],
E. M. Szanto\addressmark[USP],
A. Szanto de Toledo\addressmark[USP]
}

\runtitle{$^{6}$Li direct breakup lifetimes}
\runauthor{F. A. Souza}

\begin{document}

\maketitle

\begin{abstract}
$\alpha$-$d$ coincidence data were studied for the $^6$Li~+~$^{59}$Co 
reaction at E$_{lab}$ = 29.6~MeV. By using a kinematic analysis, it was 
possible to identify which process, leading to the same final state, has 
the major contribution for each of the selected angular regions. Contributions 
of the $^6$Li sequential and direct breakup to the incomplete fusion/transfer 
process were discussed by considering the lifetimes obtained by using a 
semiclassical approach, for both breakup components.
\end{abstract}

\section{Introduction}

The influence of projectile breakup in reactions with stable weakly bound 
nuclei on fusion cross section has been extensively investigated in the last 
decade and more recently for reactions involving exotic nuclei 
\cite{Canto06,Keeley07,Beck03,Diaz02,Beck04,Dasgupta04,Marti05,Shrivastava06,Pakou06,Beck06,Beck07,Beck07b,Beck08,Santra09,Souza09,Souza09b}. 
The light particle spectra measured in `singles' mode display significant 
contributions from reaction mechanisms other than projectile breakup. This was 
for example shown very recently for $^{6}$Li + $^{59}$Co, which can be 
considered as a benchmark
system~\cite{Beck03,Beck04,Beck06,Beck07,Beck07b,Beck08,Souza09,Souza09b}. 
Therefore, coincidence measurements are of crucial importance to disentangling 
the respective contributions of the non-capture projectile breakup components 
(both direct and sequential) from other competing mechanisms such as 
incomplete fusion (ICF) and/or transfer (TR). The contributions of sequential 
projectile  breakup (SBU) and direct projectile breakup (DBU) are both 
significant and it is necessary to determine which process influences complete 
fusion (CF) most. In this case, the study of the breakup dynamics could 
provide decisive information. 

\section{Experimental Setup}
\label{sec:expersetup}

The $^{6}$Li beam was delivered by the 8~UD Tandem accelerator at the the 
University of S\~ao Paulo Pelletron Laboratory and bombarded a 
$2.2$~mg/cm$^{2}$ $^{59}$Co target with an effective energy of $E_{lab} = 
29.6$~MeV. Eleven triple telescopes (ion chamber, a Si surface barrier 
detector and a CsI detector) \cite{Moura01} were used for the detection and 
identification of the light particles, positioned on both sides with respect 
to the beam direction with 10$^{\circ}$ angular steps, covering angular 
intervals from $-45^{\circ}$ to $-15^{\circ}$ and $15^{\circ}$ to 
$75^{\circ}$. 

\section{Results and discussion}
\label{sec:results}

The study of $\alpha$ particles and $d$ energy spectra for ``single'' mode 
events, performed in our previous work~\cite{Souza09}, indicated a dominant 
contribution of ICF/TR, more specifically $d$-incomplete fusion 
($d$-ICF)/$d$-transfer ($d$-TR) for the case of $\alpha$ spectra and 
$\alpha$-incomplete fusion ($\alpha$-ICF)/$\alpha$-transfer ($\alpha$-TR) 
for the $d$ spectra. In each case, the corresponding intermediate nucleus 
excitation energies were $24.6$~MeV and $22.5$~MeV for the $^{61}$Ni and 
$^{63}$Cu nuclei, respectively.

In this work, the contributions of the processes mentioned above, as well 
as the $^{6}$Li projectile breakup were identified for different angular 
regions selected by using the $\alpha$-$d$ coincidence measurements together 
with a 3-body kinematics analysis. The behavior of the energy centroid of 
a broad structure present in the energy spectra was studied as a function 
of the angle. For angular differences within the $^{6}$Li breakup cone 
corresponding to the (2.186~MeV, 3$^+$) first resonant state, we observe 
two sharp peaks from the two possible kinematical solutions of the SBU and 
also a broad structure is observed between them. No other resonant state was 
observed in the data. For angular differences larger than the SBU cone, we 
observed only broad structures. These broad structures could be associated 
either with the decay of nuclei produced in ICF/TR (incomplete fusion and/or 
transfer) or to $^{6}$Li DBU to the continuum.

\begin{figure}[htb]
\begin{minipage}[t]{75mm}
\centering
\includegraphics[scale=0.80]{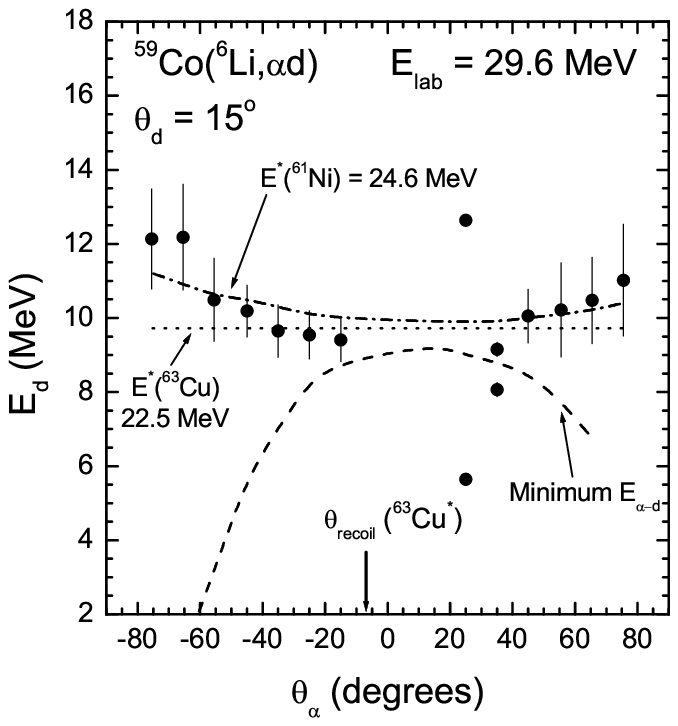}
\vspace{-5.5mm}
\caption{Experimental values for the deuteron energy as a function of the 
$\alpha$-particle detection angle. The 3-body kinematics predictions for
ICF/TR and the minimum relative energy $E_{\alpha d}$ for $^{6}$Li breakup.}
\label{fig:Energies}
\end{minipage}
\hspace{\fill}
\begin{minipage}[t]{75mm}
\centering
\includegraphics[scale=0.80]{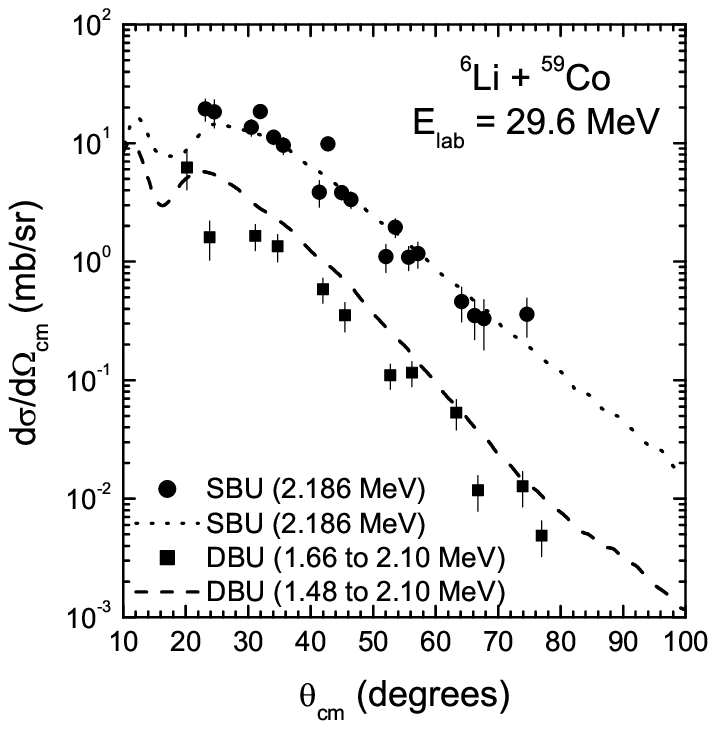}
\vspace{-5.5mm}
\caption{Experimental angular distributions for sequential (SBU - full 
circles) and direct (DBU - full squares) breakup processes. The corresponding 
CDCC calculations~\cite{Beck07,Beck07b,Beck08} are represented by dotted and 
dashed lines respectively.}
\label{fig:AngDistr}
\end{minipage}
\end{figure}

Figure~\ref{fig:Energies} shown the $d$ energy $E_{d}$ as a function of 
$\theta_{\alpha}$ for $\theta_{d}=15^{\circ}$. In this case, if 
$\alpha$-ICF/TR is dominant, the $d$ energy $E_{d}$ should be constant, 
consistent with the $22.5$~MeV excitation energy of the $^{63}$Cu 
intermediate nucleus (dotted line). This behavior would be more evident for 
angles near the $^{63}$Cu recoil direction, for which we expect the maximum 
of cross section for the $\alpha$-particle decay. This is indeed observed 
for angles near the recoiling $^{63}$Cu. For other negative angles we observe 
instead a trend consistent with a $24.6$~MeV excitation energy for the 
$^{61}$Ni composite system (dot-dashed line). This suggests the dominance 
of the $d$-ICF/TR process. Therefore, both $\alpha$-ICF/TR and $d$-ICF/TR 
contributions can be, in principle, mixed together. As shown in 
\cite{Scholz77,Mason92}, if $^{6}$Li direct breakup is dominant, the 
centroid of the broad structure would approximately correspond to the 
minimum allowed $\alpha$-$d$ relative energy ($E_{\alpha-d}$) for each 
angular pair. This trend is observed in Fig.~\ref{fig:Energies} (dashed 
line) and suggests that the $^{6}$Li DBU dominates in the case of angular 
pairs for which the broad structure is observed with 
$\Delta\theta_{\alpha d} = 10^{\circ}$ and $20^{\circ}$. For these angular 
pairs (i.e. for $\theta_{\alpha}$ angle values ranging between $20^{\circ}$ 
and $40^{\circ}$) the experimental points shown in Fig.~\ref{fig:Energies} 
correspond to the energies of the SBU peaks.

The dynamics of the SBU and DBU processes were investigated using a 
semiclassical approach, as the one previously adopted in 
Ref.~\cite{Kanungo96}. Since the Sommerfeld parameter ($\eta\sim6$) is 
sufficient large and the relative energies of the breakup fragments are not 
too high, we can assume that the projectile, as well as the breakup 
fragments, follow a Coulomb trajectory.

\begin{figure}[htb]
\begin{minipage}[t]{75mm}
\centering
\includegraphics[scale=0.80]{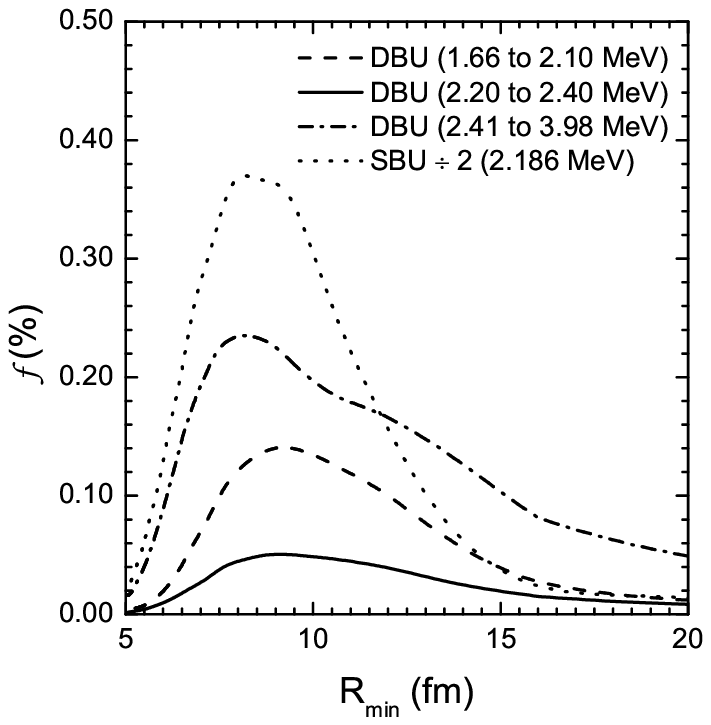}
\vspace{-5.5mm}
\caption{Function $f$ representing the probability of particle production as a function of the distance of closest approach $R_{min}$ for the SBU and DBU processes.}
\label{fig:Functionf}
\end{minipage}
\hspace{\fill}
\begin{minipage}[t]{75mm}
\centering
\includegraphics[scale=0.80]{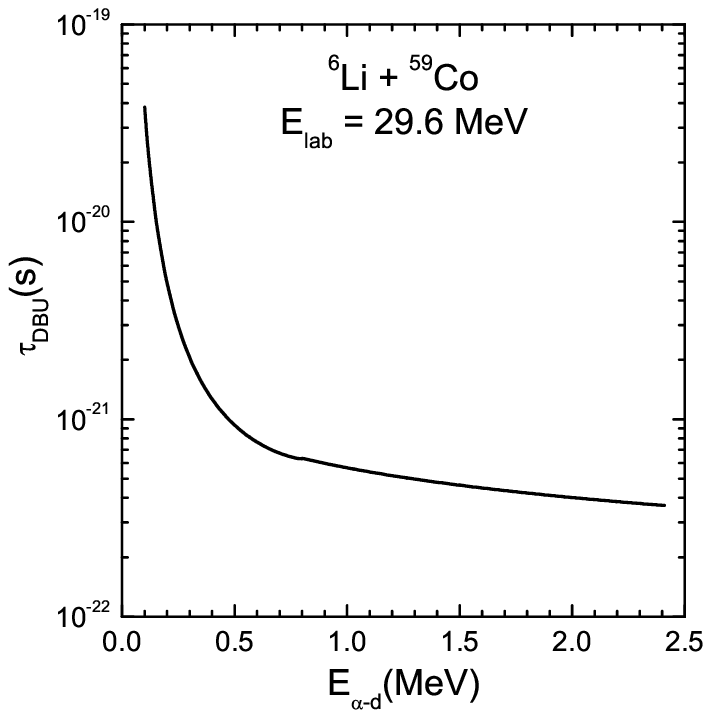}
\vspace{-5.5mm}
\caption{Calculated lifetimes of the DBU continuum states as a function of the relative 
energy $E_{\alpha d}$.}

\label{fig:lifetime}
\end{minipage}
\end{figure}

Figure~\ref{fig:AngDistr} depicts the experimental angular distribution for 
the SBU process analyzed in Ref.~\cite{Souza09}, as well as for the DBU. The 
angular distribution for the DBU is shown for $^{6}$Li continuum excitation 
energies summed between E$^*$ = $1.66$ MeV and E$^*$ = $2.10$~MeV. The dotted 
line was extracted from Ref.~\cite{Souza09} and corresponds to the SBU CDCC 
calculation \cite{Beck07,Beck07b,Beck08}. The dashed line represents the DBU 
CDCC result \cite{Beck07.Beck07b,Beck08} for a $^{6}$Li excitation energy 
range from E$^*$ =  $1.48$~MeV (breakup threshold) to E$^*$ = $2.10$~MeV. 
The curves presented in this figure were used to calculate the relative 
probability of particle production (function $f$) for SBU and DBU processes 
as a function of the distance of closest approach $R_{min}$ as shown in 
Fig.~\ref{fig:Functionf}. In the same figure we also present the function 
$f$ for the other two DBU excitation energy ranges E$^*$ =  $2.20$ to 
$2.40$~MeV and E$^*$ = $2.41$ to $3.98$~MeV not shown in 
Fig.~\ref{fig:AngDistr}. In the semiclassical calculations we assumed the 
most probable value of the excitation energy observed experimentally for 
each energy range ($E^{*} = 1.7, 2.3$ and $3.2$~MeV).

The DBU lifetimes ($\tau_{DBU}$) due to barrier tunneling were estimated 
adopting the model of Ref.~\cite{Tokimoto01} and the calculated 
$\tau_{DBU}$ as a function of $E_{\alpha d}$ is presented in 
Fig.~\ref{fig:lifetime}.

From Fig.~\ref{fig:Functionf} one can notice that the most probable value of 
$R_{min}$ are very similar for SBU and DBU. However, for the $^{6}$Li $3^{+}$ 
state with $E^{*} = 2.186$~MeV the SBU lifetime is $\tau_{SBU} = 2.73 \times 
10^{-20}$~s corresponding to $\Gamma_{SBU} = (0.024 \pm 0.002)$~MeV 
\cite{Tilley02}, which is at least one order of magnitude greater than the 
ones observed in Fig.~\ref{fig:lifetime} for DBU. It indicates that  
sequential projectile breakup occurs very far from the target. On the other 
hand, for DBU the shorter lifetimes of the continuum `states' cause the \
breakup process to occur at shorter distances from the target. Therefore, in 
reactions that have a large ICF/TR cross section (as 
Refs.~\cite{Souza09,Santra09}, for instance) and a major influence on the 
complete fusion cross section, the flux diverted from CF to ICF/TR would 
arise essentially from the DBU components (higher excitation energies in 
the continuum).


\begin{thebibliography}{99}

\bibitem{Canto06} L.F. Canto, P.R.S. Gomes, R. Donangelo, and M.S. Hussein, 
Phys. Rep. \textbf{424}, 1 (2006) and references therein.

\bibitem{Keeley07} N. Keeley, R. Raabe, N. Alamanos, and J.L. Sida, 
Prog. Part. Nucl. Phys. \textbf{59}, 579 (2007); 
arXiv: \textbf{nucl-th/0702038} (2007) and references therein.

\bibitem{Beck03} C. Beck \textit{et al.}, Phys. Rev. C \textbf{67}, 054602 
(2003).

\bibitem{Diaz02} A. Diaz-Torres, I.J. Thompson, Phys. Rev. C \textbf{65}, 
024606 (2002); arXiv:\bf nucl-th/0111051 \rm (2001). 

\bibitem{Beck04} C. Beck \it et al.,\rm arXiv:\bf nucl-ex/0411002 \rm (2004)

\bibitem{Dasgupta04} M. Dasgupta \textit{et al.}, Phys. Rev. C \textbf{70}, 
024606 (2004).

\bibitem{Marti05} G. V. Marti \textit{et al.}, Phys. Rev. C \textbf{71}, 027602 
(2005).

\bibitem{Shrivastava06} A. Shrivastava \textit{et al.}, Phys Lett. B 
\textbf{633}, 463 (2006); \rm arXiv:\bf nucl-ex/0512032 \rm (2005).

\bibitem{Pakou06} A. Pakou \textit{et al.}, Phys. Lett. B \textbf{633}, 691 (2006).

\bibitem{Beck06} C. Beck, A. S\`anchez i Zafra, A. Diaz-Torres, I.J. Thompson, 
N. Keeley, and F.A. Souza, AIP Conferences Proceedings \textbf{853}, 384 
(2006); Proc. of Fusion06 Conference, San Servolo, Venezia, Italy, 19-23 March 
2006; \rm arXiv:\bf nucl-th/0605029 \rm (2006).
 
\bibitem{Beck07} C. Beck, N. Keeley and A. Diaz-Torres, Phys. Rev. C \textbf{75}, 
054605 (2007); \rm arXiv:\bf nucl-th/0703085 \rm (2007).

\bibitem{Beck07b} C. Beck, Nucl. Phys. A \textbf{787}, 251 (2007);
  \rm arXiv:\bf nucl-ex/0701073 \rm (2007);
  \rm arXiv:\bf nucl-th/0610004 \rm (2006).

\bibitem{Beck08} C. Beck, N. Keeley, and A. Diaz-Torres, \rm  AIP Conferences 
Proceedings \bf 1012\rm, 233 (2008); Proc. of Frontiers in Nuclear Structure,
Astrophysics, and Reactions (FINUSTAR2), Crete, Greece, 10-14 September 2007; 
\rm arXiv: \bf 0709.0439 \rm (2008). 
  
\bibitem{Santra09} S. Santra \textit{et al.}, \rm Phys. Lett. B \textbf{677}, 139 
(2009).

\bibitem{Souza09} F. A. Souza \textit{et al.}, Nucl. Phys. A \textbf{821}, 36 
(2009); \rm arXiv:\bf 0811.4556 \rm (2009).

\bibitem{Souza09b} F.A. Souza, N. Carlin, C. Beck, N. Keele, A. Diaz-Torres, 
R. Liguori Neto, C. Siqueira-Mello, M.M. de Moura, M.G. Munhoz, R.A.N. 
Oliveira, M.G. Del Santo, A.A.P. Suaide, E.M. Szanto, and A. Szanto de 
Toledo, submitted to Eur. Phys. Journal A; \rm arXiv: \bf 0909.5556 \rm (2009). 

\bibitem{Moura01} M. M. de Moura \textit{et al.}, Nucl. Instr. Meth. A 
\textbf{471}, 368 (2001).

\bibitem{Scholz77} D. Scholz \textit{et al.}, Nucl. Phys. A \textbf{288}, 351 
(1977).

\bibitem{Mason92} J. E. Mason \textit{et al.}, Phys. Rev. C \textbf{45}, 2870 
(1992).

\bibitem{Kanungo96} R. Kanungo \textit{et al.}, Nucl. Phys. A \textbf{599}, 579 
(1996).

\bibitem{Tokimoto01} Y. Tokimoto \textit{et al.}, Phys. Rev. C \textbf{63}, 
035801 (2001).

\bibitem{Tilley02} D.R. Tilley \textit{et al.}, Nucl. Phys. A \textbf{708}, 3 
(2002).

\end{thebibliography}
\end{document}